\documentclass[11pt]{article}
\pdfoutput=1
\usepackage{%
amssymb,%
graphicx,%
epstopdf,%
amsmath,%
amsfonts,%
epsfig,%
graphics,%
xcolor,%
soul
}

\begin{document}

\begin{flushright}
  INR-TH-2020-029
\end{flushright}

\sloppy

\title{
\bf
Gravitational reheating and superheavy Dark Matter creation after inflation with non-minimal coupling
}
\author{E.~Babichev$^{a}$, D.~Gorbunov$^{b,c}$, S.~Ramazanov$^{d}$, L.~Reverberi$^{d}$\\
\small{$^a$\em Universit\'e Paris-Saclay, CNRS/IN2P3, IJCLab, 91405 Orsay, France,}\\
\small{$^b$\em Institute for Nuclear Research of the Russian Academy of Sciences,}\\
\small{\em 60th October Anniversary prospect 7a, Moscow 117312, Russia}\\
\small{$^c$\em Moscow Institute of Physics and Technology, Institutsky per. 9, Dolgoprudny 141700, Russia}\\
\small{$^d$\em CEICO, Institute of Physics of the Czech Academy of Sciences,}\\
\small{\em Na Slovance 1999/2, 182 21 Prague 8, Czech Republic}
}

{\let\newpage\relax\maketitle}

\begin{abstract}

  We discuss the gravitational creation of superheavy particles $\chi$ in an inflationary scenario with a quartic potential and a non-minimal coupling between the inflaton $\varphi$ and the Ricci curvature: $\xi \varphi^2 R/2$.
  We show that for large constants $\xi \gg 1$, there can be abundant production of particles $\chi$ with masses largely exceeding the inflationary Hubble rate $H_{\rm infl}$, up to $(\text{a few}) \times \xi H_{\rm infl}$, even if they are conformally coupled to gravity. We discuss two scenarios involving these gravitationally produced particles $\chi$. In the first scenario, the inflaton has only gravitational interactions with the matter sector and the particles $\chi$ reheat the Universe. 
In this picture, the inflaton decays only due to the cosmic expansion, and effectively contributes to dark radiation, which can be of the observable size. The existing limits on dark radiation lead to an upper bound on the reheating temperature.
  In the second scenario, the particles $\chi$ constitute Dark Matter, if substantially stable. In this case, their typical masses should be in the ballpark of the Grand Unification scale.

\end{abstract}

\section{Introduction}
\label{section_introduction}

It is well-known that particles can be produced in curved space-times, even if they have only gravitational interactions~\cite{Zeldovich:1971mw}. For example, creation of particles takes place in Friedmann-Lema\^itre-Robertson-Walker (FLRW) background due to the changing scale factor $a(t)$~\cite{Parker:1969au, Grib:1970xx}.

In cosmology, one typically assumes that gravitational particle production is efficient only for masses not largely exceeding the Hubble rate $H \equiv \dot a/a$ at the end of inflation; for heavier masses an exponential suppression comes into play~\cite{Kuzmin:1998uv, Kuzmin:1998kk, Chung:1998ua, Chung:1998zb}. This is indeed the case when the Hubble rate changes on the time scales $\sim H^{-1}$ or slower. Such a cosmological evolution is common in simple scenarios with a canonical inflaton minimally coupled to gravity. However, there are models predicting a rapidly changing Hubble rate in the post-inflationary Universe. In that case, the masses of the produced particles can be considerably larger than the inflationary Hubble rate~\cite{Ema:2018ucl, Chung:2018ayg, Ema:2019yrd}.

In the present work, we focus on the latter type of models, and in particular on ones involving an inflaton $\varphi$ equipped with a quartic self-interaction and a non-minimal coupling to the Ricci curvature, i.e., $\xi \varphi^2 R/2$. In this class, Higgs inflation is perhaps the most notable example~\cite{Bezrukov:2007ep}.
However, we will discuss the models of interest from a broader prospective. From the viewpoint of particle production, the key feature of these scenarios is the presence of spikes in the post-inflationary evolution of the inflaton and Hubble rate time derivatives~\cite{Ema:2016dny}. These spikes appear shortly after the end of inflation around the zero-crossings of the inflaton and get smoother with time. The time scale of the first spikes is estimated as $(\xi \cdot H_{\rm infl})^{-1}$, where $H_{\rm infl}$ is the characteristic inflationary Hubble rate (see Ref.~\cite{Ema:2016dny} and the discussion in Section~\ref{section_evolution_inflaton_Hubble}). Hence, for $\xi \gg 1$, the inflaton and the Hubble rate change very rapidly during a Hubble time.

This opens up the opportunity of efficient gravitational production of super-Hubble particles $\chi$ with masses up to $m_{\chi} \simeq \xi H_{\rm infl}$. We compute the energy density of these particles assuming a conformal coupling to gravity in Section~\ref{section_gravitational_production}. To achieve this, we find analytical expressions for the inflaton and the Hubble rate in the vicinity
of the first spike. Using these, we calculate analytically the Bogolyubov coefficient, which defines the number density of particles $\chi$. We compare our analytical expressions with the results of numerical calculations, and find an excellent agreement.
We show that for masses $m_{\chi} \lesssim \xi H_{\rm infl}$, particles are indeed produced 
with no exponential suppression. Consequently, the particles $\chi$ may constitute a considerable fraction of the energy budget of the Universe, and leave potentially interesting imprints in the cosmological evolution.

In this work, we consider two scenarios involving the particles $\chi$. In the first scenario they reheat the Universe in the situation in which the inflaton interacts only gravitationally with other matter fields (Section~\ref{section_reheating}).
The energy density of the non-relativistic $\chi$-particles is large enough to dominate over the inflaton energy density, which redshifts as $1/a^4$, after some time. Later on, the $\chi$-particles decay into Standard Model (SM) species, and the Universe gets reheated.
The key prediction of this scenario is effective dark radiation in the form of the inflaton condensate,
which decays only due to the Hubble drag. Dark radiation can be constrained through
the measurements of the effective number of neutrino species. This yields a mass-dependent upper bound on the reheating temperature, which turns out to be relatively low.
In particular, for masses slightly above $\xi H_{\rm infl}$, reheating in this scenario is in conflict with the requirement of successful Big Bang nucleosynthesis (BBN). Yet there is another advantage of our reheating mechanism (besides predictability): it avoids the problem of overproduction of gravitational waves common for the simplest models of gravitational reheating~\cite{Figueroa:2018twl} (see also Ref.~\cite{Artymowski:2017pua}), where the inflaton decays only because of the cosmic expansion.
Note also that the absence of direct interactions of the inflaton with matter fields guarantees the flatness of the inflationary potential, which would receive possibly large quantum corrections otherwise.
In other words, one can trust the inflationary predictions derived in the single-field approach.

The second scenario deals with applications of $\chi$-particles for DM. The obstacle here is that for not extremely large masses, $m_{\chi} \ll \xi H_{\rm infl}$, the energy density of $\chi$-particles is well above the required DM abundance.
A way out of this problem is to assume that the particles $\chi$ are unstable and the DM particles appear as their decay products (Subsection~\ref{section_decay_products}). On the other hand, for $m_{\chi} \gtrsim \xi H_{\rm infl}$, particle creation is exponentially suppressed. Therefore, DM composed of $\chi$-particles stable on time scales much larger than the present age of the Universe can be produced with the right abundance (Subsection~\ref{section_chi_as_dark_matter}).

The paper is organized as follows. In Section~\ref{section_model}, we introduce the main ingredients of the model. In Section~\ref{section_evolution_inflaton_Hubble}, we discuss the post-inflationary evolution of the inflaton and the Hubble rate in inflation with a non-minimal coupling.
Then, we compute analytically and numerically production of superheavy particles in Section~\ref{section_gravitational_production}.
In Section~\ref{section_reheating}, we discuss the reheating scenario involving the produced particles. We consider particles $\chi$ as DM in Section~\ref{section_dark_matter}, and conclude in Section~\ref{section_summary}.

\section{The model}
\label{section_model}

We consider the action given by
\begin{equation}
  \label{modelbasic}
  S=S_{EH}+S_{\rm infl}+S_{\chi}+S_{SM}+S_{int} \; .
\end{equation}
Here $S_{EH}$ is the Einstein-Hilbert action:
\begin{equation}
  \nonumber
  S_{EH}=-\frac{M^2_{Pl}}{2} \int d^4 x \sqrt{-g} R \; .
\end{equation}
We use a mostly negative signature of the metric; $M_{Pl} \approx 2.44 \cdot 10^{18}~\mbox{GeV}$ is the reduced Planck mass. The term $S_{\rm infl}$ describes the action of an inflaton $\varphi$, which we assume to be non-minimally coupled to gravity:
\begin{equation}
  \label{inflaction}
  S_{\rm infl}= \int d^4 x \sqrt{-g} \left[\frac{\left(\partial_{\mu} \varphi \right)^2}{2} -\frac{\lambda \varphi^4}{4} -\frac{\xi \varphi^2 R}{2} \right] \; .
\end{equation}
Hereafter, we work in the Jordan frame. The model~(\ref{inflaction}) was studied in the context of Higgs inflation~\cite{Bezrukov:2007ep}, where the SM scalar plays the role of the field $\varphi$. However, we assume a more generic setup, where the field $\varphi$ has a different nature.

The action $S_{\chi}$ describes the dynamics of the superheavy field $\chi$, a singlet scalar also non-minimally coupled to gravity:
\begin{equation}
  \label{chiaction}
  S_{\chi}=\int d^4 x \sqrt{-g} \left[\frac{(\partial_{\mu} \chi)^2}{2}-\frac{m^2_{\chi} \chi^2}{2}+\frac{\zeta \chi^2 R}{2} \right] \; .
\end{equation}
Here $m_{\chi}$ is the mass of the superheavy field; the coupling constant to the Ricci curvature $\zeta$ will be specified later. The action $S_{SM}$ in Eq.~(\ref{modelbasic}) takes into account SM fields plus possibly sterile neutrinos responsible for the small masses of active neutrinos.
Finally, $S_{int}$ contains interactions between the inflaton $\varphi$, the superheavy field $\chi$, and SM fields. The form of $S_{int}$ will be made explicit when relevant. In all the scenarios considered in this paper, we assume that the superheavy field $\chi$ is only gravitationally coupled to the inflaton $\varphi$ and has at most very weak couplings to the SM fields.

\section{Evolution of inflaton and Hubble rate}
\label{section_evolution_inflaton_Hubble}
First, let us discuss the cosmological evolution when the inflaton $\varphi$ gives the dominant contribution to the total energy density of the Universe.
The modified (due to the non-minimal coupling of the inflaton to gravity) Friedmann equation is given by
\begin{equation}
  \label{Fried}
  3 \left(M^2_{Pl} H^2+\xi \varphi^2 H^2 +2\xi \varphi \dot{\varphi} H \right)=\frac{1}{2}\dot{\varphi}^2+\frac{\lambda}{4} \varphi^4 \; .
\end{equation}
The background equation of motion for the inflaton reads
\begin{equation}
  \label{infleq}
  \ddot{\varphi}+3H \dot{\varphi}+\lambda \varphi^3 -6 \xi \varphi \left( 2H^2 +\dot{H} \right)=0 \; .
\end{equation}
We do not consider inflationary perturbations in the present work. However, it is well-known that their evolution is in an excellent agreement with the Cosmic Microwave Background measurements (CMB)~\cite{Akrami:2018odb}, provided that the following constraint is imposed:
\begin{equation}
  \label{constraint}
  \xi \approx 49000 \cdot \sqrt{\lambda} \; .
\end{equation}
This relation holds for sufficiently large $\xi \gtrsim 0.1$~\cite{Bezrukov:2013fca}. In particular, for Higgs inflation $\lambda \simeq 0.1$ and $\xi \simeq 10^{4}$. As we have mentioned before, we do not assume Higgs inflation in the present work. This allows us to consider much smaller values of $\xi$ (still $\xi \gg 1$).

\begin{figure}[tb!]
  \begin{center}
    \includegraphics[width=0.48\columnwidth,angle=0]{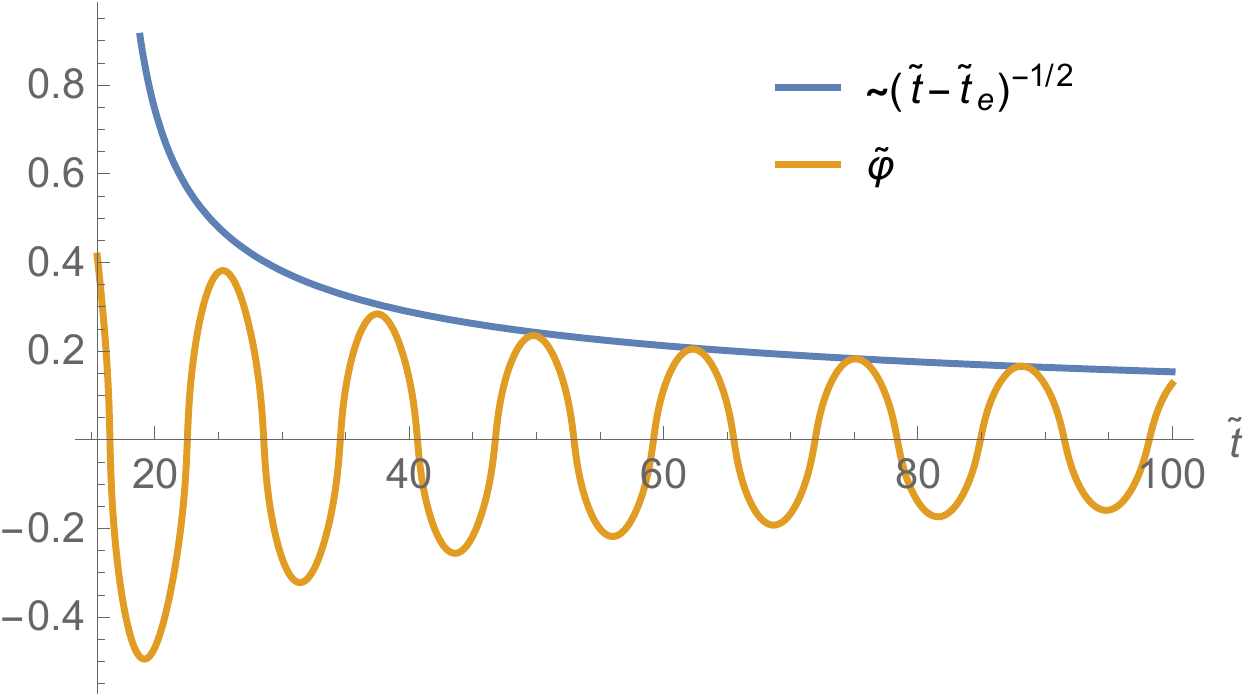}
    \includegraphics[width=0.48\columnwidth,angle=0]{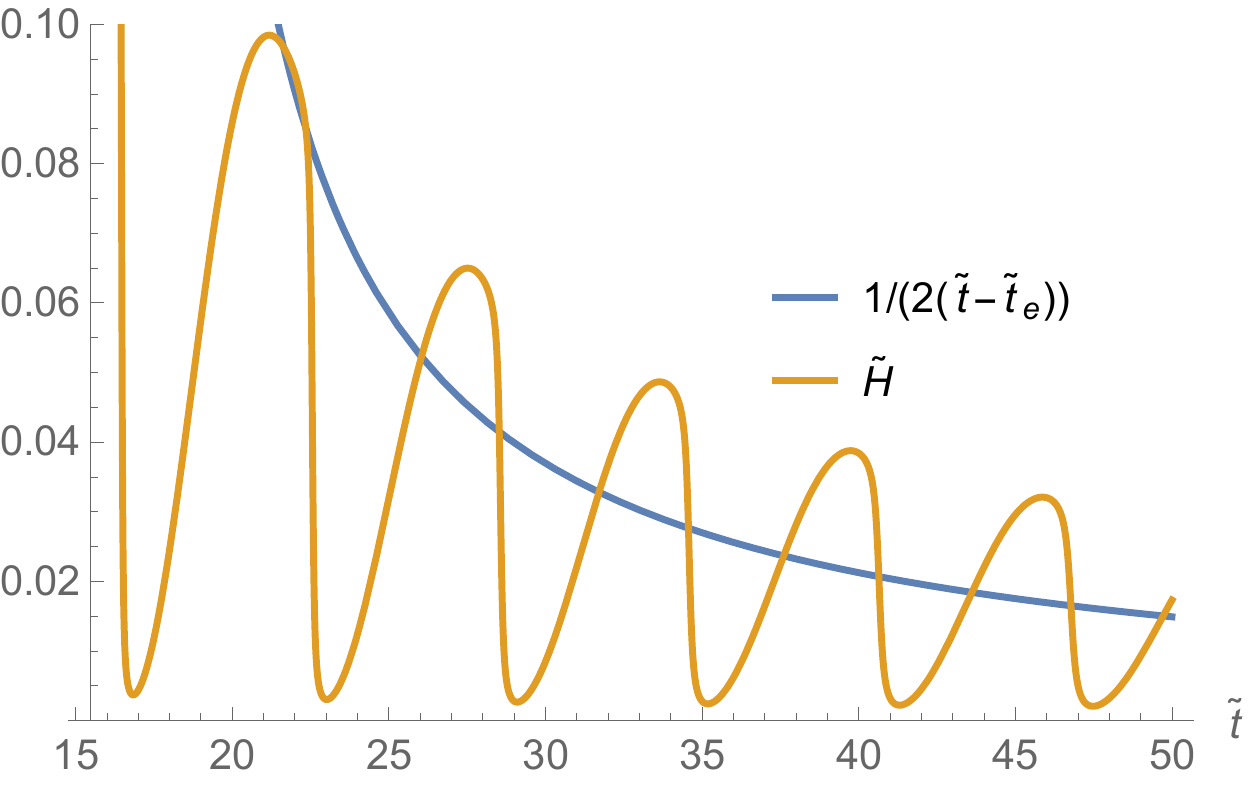}
    \caption{Dynamics of the inflaton (left) and the Hubble rate (right) in the post-inflationary Universe is shown for the model~\eqref{inflaction}, with $\xi=50$. We use the dimensionless variables of Eq.~\eqref{dimensionless}. The time $\tilde t_e$ denotes the end of inflation. The average cosmological evolution driven by the inflaton mimics that of a Universe filled with radiation: the Hubble rate approaches the behaviour ${\tilde H=1/(2\tilde t)}$ as $\tilde t$ increases.}
    \label{fig_inflatonHubble}
  \end{center}
\end{figure}

\begin{figure}[tb!]
  \begin{center}
    \includegraphics[width=0.48\columnwidth,angle=0]{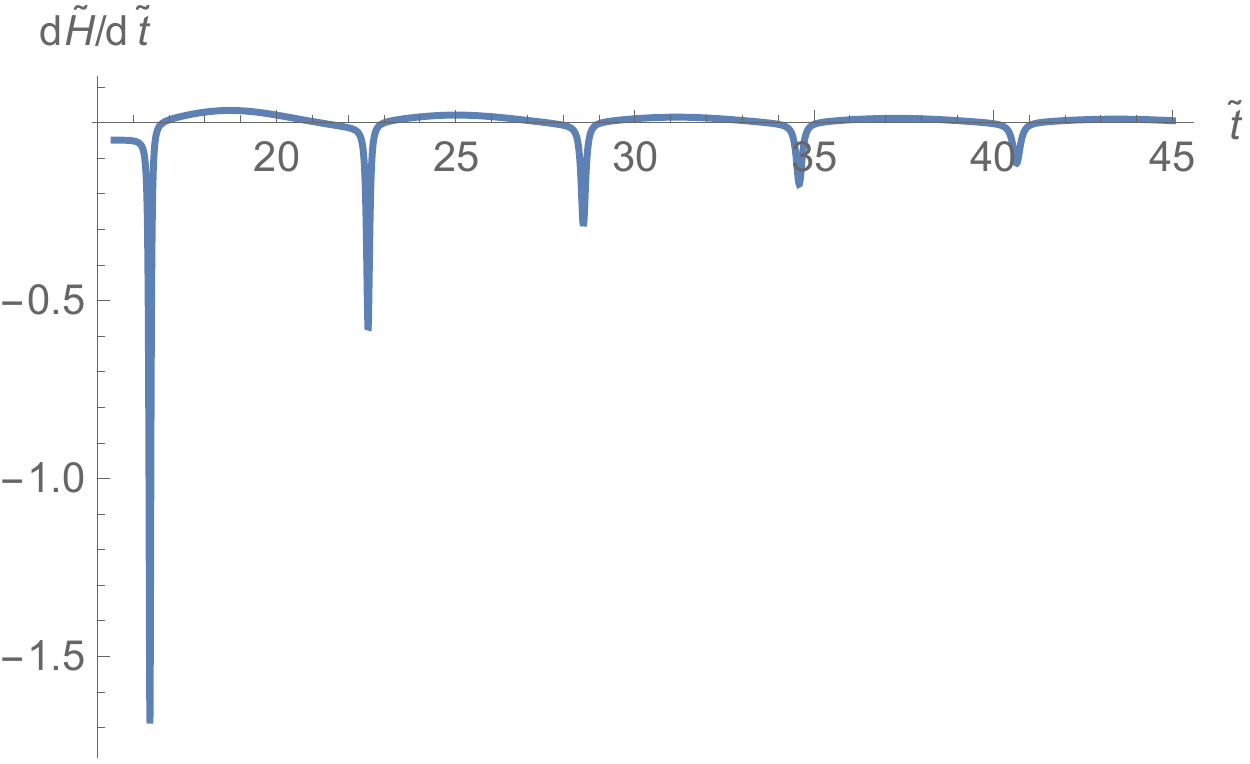}
    \includegraphics[width=0.48\columnwidth,angle=0]{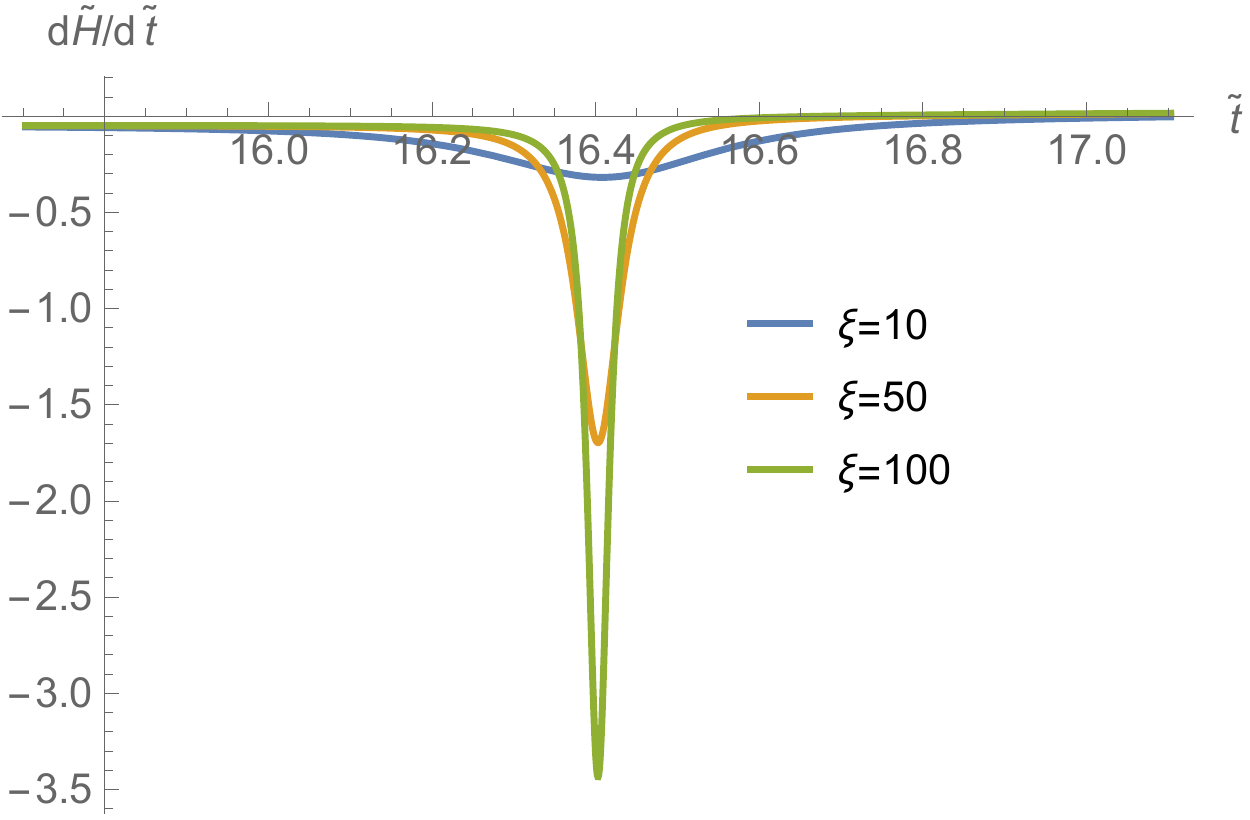}
    \caption{Evolution of the time derivative of the Hubble rate in the post-inflationary Universe is shown using the dimensionless variables of Eq.~\eqref{dimensionless}.
    The non-minimal coupling of the inflaton to gravity leads to the appearance of spikes in the evolution of $\dot{\tilde{H}}$ (left panel, $\xi=50$). The height and the width of the spikes depend on the coupling constant $\xi$, as it is seen on the right panel, where the first spike is plotted for $\xi=10, 50, 100$.}
    \label{fig_Hubble}
  \end{center}
\end{figure}

A detailed study of the post-inflationary evolution of the inflaton and the Hubble rate was performed in Ref.~\cite{Ema:2016dny}; see also Refs.~\cite{DeCross:2015uza, DeCross:2016fdz, vandeVis:2020qcp} for an extension to the multifield case. In this Section, we summarize our results of independent analytical and numerical analyses; details of the analytical calculations can be found in Appendix~\ref{appendix_details_post_inflat}.
For the purpose of numerical calculations, it is convenient to rewrite equations~\eqref{Fried} and~\eqref{infleq} in terms of the dimensionless variables
\begin{equation}
  \label{dimensionless}
  \tilde{H}=\frac{H}{H_{\rm infl}} \qquad \tilde{\varphi}=\frac{\sqrt{\xi} \varphi}{M_{Pl}} \qquad \tilde{t}=H_{\rm infl} t \; ,
\end{equation}
where we defined
\begin{equation}
  \label{hubbledef}
  H_{\rm infl} \equiv \frac{\sqrt{\lambda} M_{Pl}}{\xi} \approx 5 \cdot 10^{13}~\mbox{GeV} \; ,
\end{equation}
which corresponds to the value of the Hubble rate approximately $10$ e-foldings before the end of inflation. Then, the dimensionless version of Eqs.~(\ref{Fried}) and (\ref{infleq}) reads
\begin{equation}
  \label{friedmanndim}
  3 \left(\tilde{H}^2 +\tilde{\varphi}^2 \tilde{H}^2 +2 \tilde{\varphi} \dot{\tilde{\varphi}} \tilde{H} \right)=\frac{\dot{\tilde{\varphi}}^2}{2\xi} +\frac{\tilde{\varphi}^4}{4}
\end{equation}
and
\begin{equation}
  \label{inflatondim}
  \ddot{\tilde{\varphi}} + 3\tilde{H} \dot{\tilde{\varphi}} +\xi \tilde{\varphi}^3-6\xi \tilde{\varphi} \left(2\tilde{H}^2+\dot{\tilde{H}} \right) =0 \; .
\end{equation}
(We keep using the dot notation for the derivative with respect to dimensionless time $\tilde{t}$). The results of numerical calculations are shown in Figs.~\ref{fig_inflatonHubble} and~\ref{fig_Hubble}. The former shows the evolution of the inflaton and the Hubble rate in the post-inflationary Universe. We found that the value of the Hubble rate at the end of inflation
is in a good agreement with our analytical estimate in Appendix~\ref{appendix_details_post_inflat}:
\begin{equation}
  \label{hubbleend}
  H_e \simeq \frac{1}{6} H_{\rm infl} \; .
\end{equation}
More important for our discussion is the fact that the evolution of the time derivative of the Hubble rate $\dot{H}$ has interesting features. In Fig.~\ref{fig_Hubble}, one clearly sees the spikes of $\dot{H}$, centred around the zero-crossings of the inflaton and having a short duration $\Delta t_{\rm spikes} \ll H^{-1}$. Note that the presence of spikes is crucial to understand (p)reheating in inflation with a non-minimal coupling to gravity~\cite{Ema:2016dny}. The spikes are also present in the Einstein frame~\cite{Ema:2016dny}. 
However, they are manifested in a different way, namely, in a very fast change of the shape of the inflaton potential 
at the end of inflation.

Notably, in the vicinity of the spikes, the system of equations~\eqref{friedmanndim} and~\eqref{inflatondim} can be solved analytically. The solutions for the inflaton and the Hubble rate are derived in Appendix~\ref{appendix_details_post_inflat}. In particular, the width of the first peak is found to be (see Eq.~\eqref{widthapp})
\begin{equation}
  \label{halfwidth}
  \Delta t_1 \simeq \frac{5}{\xi H_{\rm infl}} \; ,
\end{equation}
which fits well the behaviour of spikes on Fig.~\ref{fig_Hubble}.
We see that for $\xi \gg 1$, a new energy scale $\xi H_{\rm infl}$ appears, in agreement with the results of Ref.~\cite{Ema:2016dny}. We will also need the height of the first peak of $\dot H$, which is found in Appendix~\ref{appendix_details_post_inflat}, see Eq.~\eqref{analyticalfinal},
\begin{equation}
  \label{hubblederivative}
  \left|\dot{H} (t_1) \right | \simeq \frac{\xi}{24} \cdot H^2_{\rm infl} \; .
\end{equation}
There is only a marginal disagreement between the above analytical estimate and the numerical results, namely a factor $1/30$ instead of $1/24$ in Eq.~(\ref{hubblederivative}). In the following we ignore the difference between these two factors.

One important comment is in order here. The strong coupling scale in the model~\eqref{inflaction} is given by~\cite{Burgess:2009ea, Barbon:2009ya}
\begin{equation}
  \label{strong}
  \Lambda_{str} \sim \frac{M_{Pl}}{\xi} \; .
\end{equation}
Hence, the energy scale $(\Delta t_1)^{-1}$ can be trusted only for $\xi$ limited by
\begin{equation}
  \label{nostrong}
  \xi \lesssim \sqrt{\frac{5 M_{Pl}}{H_{\rm infl}}} \simeq 500 \; .
\end{equation}
The upper bound here corresponds to $\Lambda_{str}$ of the order of the Grand Unification scale $\Lambda_{GUT}$, i.e., $\Lambda_{str} \sim \Lambda_{GUT} \sim 10^{15}-10^{16}~\mbox{GeV}$. In addition to this, the requirement that we consistently work in the weak coupling regime leads to another condition: $\left|\dot{H} (t_1) \right | \ll \Lambda_{str}^2$. One can check that the latter is satisfied automatically provided that the constraint~\eqref{nostrong} is fulfilled. The bound~\eqref{nostrong} implies that our analysis is not applicable to Higgs inflation, for which $\xi \simeq 10^4$. For such large $\xi$, one can avoid the strong coupling problem by adding an $R^2$-term~\cite{Ema:2017rqn, Gorbunov:2018llf}. In this case, the strong coupling scale is shifted up to the Planck mass, while the spikes become smoother. We proceed assuming that the bound~\eqref{nostrong} is satisfied.

\section{Gravitational production of superheavy particles}
\label{section_gravitational_production}

Now let us study the production of $\chi$-particles described by the action~\eqref{chiaction} in the cosmological background discussed in the previous Section. Below we show that for large $\xi$, the gravitational production of the field $\chi$ is very efficient for masses up to $m_{\chi} \sim \xi H_{\rm infl}$. The gravitational production is quantified by the Bogolyubov coefficient $\beta_k$.
Assuming $|\omega'_k/\omega^2_k| \ll 1$, the latter is given by~\cite{Winitzki:2005rw}\footnote{See also Appendix in the draft of the book~\cite{Mukhanov}.}
\begin{equation}
  \label{Bogad}
  \beta_k =\int^{+\infty}_{\eta_{Pl}} d \eta \cdot \frac{\omega'_k}{2\omega_k} \cdot \mbox{exp} \left[-2i \int^{\eta}_{\eta_{Pl}} d\eta' \, \omega_k \right] \; ,
\end{equation}
where $\omega_k$ is the frequency of the mode with conformal momentum $k$:
\begin{equation}
  \label{frequency}
  \omega_k=\sqrt{k^2+a^2 m^2_{\chi}+\frac{1}{6}(1-6\zeta)a^2 R} \; .
\end{equation}
Here $\eta$ is the conformal time. The Planckian time $\eta_{Pl}$ formally corresponds to the beginning of inflation, but in practice the value of the lower limit of the integral makes no difference as long as it corresponds to early enough times, that is when there are no $\chi$-particles yet.
Note also that the choice of $\eta_{Pl}$ in the exponent is arbitrary: its effect is to change an irrelevant overall phase factor of the Bogolyubov coefficient. Substituting Eq.~\eqref{frequency} into Eq.~\eqref{Bogad} and switching to the physical time, one obtains
\begin{equation}
  \label{Bogreal}
  \beta_k =\frac{1}{2} \int^{+\infty}_{t_{Pl}}dt \left[\frac{m^2_{\chi} H(t)+\frac{1}{6} (1-6\zeta)\left(\frac{1}{2} \dot{R} (t)+H(t) R(t) \right)}{(\omega_k (t)/a(t))^2} \right] \cdot \mbox{exp} \left[-2i \int^{t}_{t_{Pl}} dt' \, \frac{\omega_k (t')}{a(t')} \right] \; .
\end{equation}
Hereafter, unless specified otherwise, we focus on the case of conformal coupling to gravity $\zeta=1/6$.
In this case, the expression above is considerably simplified:
\begin{equation}
  \label{Bogolyubovconformal}
  \beta_k =\frac{1}{2} \int^{+\infty}_{t_{Pl}} dt \cdot \left[\frac{m^2_{\chi} H(t)}{(\omega_k (t)/a(t))^2} \right] \cdot \mbox{exp} \left[-2i \int^{t}_{t_{Pl}} dt' \, \frac{\omega_k (t')}{a(t')} \right] \; .
\end{equation}
From the analytical calculations in Appendix~\ref{appendix_details_post_inflat} and the numerical calculations we can see that the derivative of the Hubble rate peaks strongly around the point in which the inflaton passes through zero, see Fig.~\ref{fig_Hubble} and Eq.~\eqref{hubblederivative}.
These spikes give the main contribution to the Bogolubov coefficient. To estimate this contribution, it is convenient to perform an integration by parts:
\begin{equation}
  \label{intermediate}
  \beta_k =-\frac{i}{4} \int^{+\infty}_{t_{Pl}} dt \, \frac{m^2_{\chi}}{(\omega_k (t)/a(t))^3} \cdot \left[\dot{H} (t)+\frac{3k^2 \cdot H^2(t) }{\omega^2_k (t)} \right]  \cdot \mbox{exp} \left[-2i \int^{t}_{t_{Pl}} dt' \, \frac{\omega_k (t')}{a(t')} \right] \; ,
\end{equation}
where the boundary terms vanish, because $a(t_{Pl}) \rightarrow 0$ and $H(\infty) \rightarrow 0$. So far, we did not make any approximations (apart from $|\omega'_k/\omega^2_k| \ll 1$ or, equivalently, $|\beta_k| \ll 1$).
Now, let us neglect the second term in the square brackets.
This is well justified, because $|\dot{H}| \gg H^2$ at the spike for $\xi\gg 1$, see Eq.~(\ref{hubblederivative}).
Furthermore we are interested in the modes $k/a(t_1) \lesssim m_{\chi}$, since modes with larger momenta give a sub-dominant contribution to the particle density, as it will become clear shortly. Hence, from Eq.~(\ref{intermediate}) we obtain
\begin{equation}
  \label{Bogolyubovapprox}
  \beta_k \approx -\frac{i}{4} \int^{+\infty}_{t_{Pl}} dt \,\frac{m^2_{\chi}\dot{H} (t)}{(\omega_k (t)/a(t))^3} \cdot \mbox{exp} \left[-2i \int^{t}_{t_{Pl}} dt'\, \frac{\omega_k (t')}{a(t')} \right] \; .
\end{equation}
The common lore is that this integral is exponentially suppressed for super-Hubble particles with masses $m_{\chi} \gg H_e$, where the subscript `$e$' denotes the end of inflation.
This is true when the Hubble rate $H$ changes on the time scale $H^{-1}$ (or slower) during the whole evolution.
In our case, however, the change of the Hubble rate at spikes is faster than $H^{-1}$. Therefore, the naive conclusion about an exponential suppression starting at the scale $H_e$ is not valid. Indeed, the width of the first spike is given by Eq.~\eqref{halfwidth}. This indicates that particles with masses up to $m_{\chi} \simeq (\Delta t_1)^{-1} \simeq \xi H_{\rm infl}/5$ can be produced without the exponential suppression, as we confirm below. 
The exponential suppression becomes efficient for heavier particles.

\begin{figure}[tb!]
  \begin{center}
    \includegraphics[width=0.7\columnwidth,angle=0]{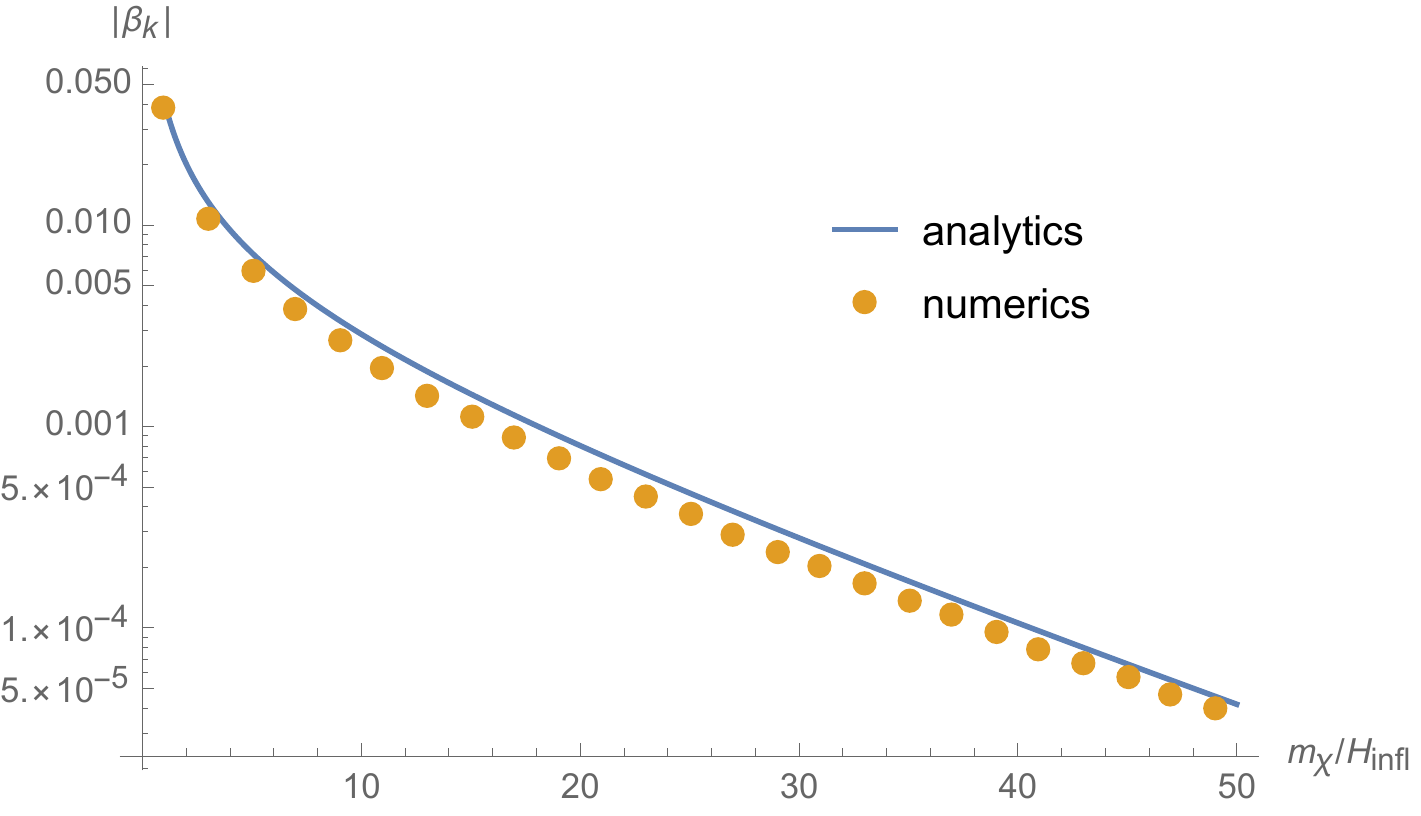}
    \caption{The absolute value of the Bogolyubov coefficient, $|\beta_k|$, is shown as the function of the mass $m_{\chi}$ for $k=0$. The constant $\xi$ is set to $\xi=50$.
    We define $H_{\rm infl}$ in Eq.~\eqref{hubbledef}. The results of numerical computation shown by the orange points are well fitted by the formula~\eqref{numericbog}, which matches
    analytically derived expression~\eqref{Bogolyubovanalytic} for the Bogolyubov coefficient (blue line).}\label{Bogolyubov}
  \end{center}
\end{figure}

We derive both analytical and numerical solutions of the Bogolyubov coefficient $\beta_k$ considering only the contribution of the first spike (see the comment below). We also neglect the time variation of the scale factor $a(t)$ in the vicinity of the spike, i.e., we replace $a(t)$ by $a_e$.
The details of analytical and numerical calculation of $\beta_k$ based on Eq.~\eqref{Bogolyubovapprox} can be found in Appendices~\ref{appendix_analytic_bogolyubov} and~\ref{appendix_numerical_bogolyubov}, respectively.
We checked that Eq.~\eqref{Bogolyubovanalytic} derived in Appendix~\ref{appendix_analytic_bogolyubov},
\begin{equation}
  \label{numericbog}
  |\beta_k| \simeq \frac{m^2_{\chi}}{6 \xi(\omega_k (t_e)/a_e)^2 }\, K_1 \left(\frac{4\cdot \omega_k (t_e)}{a_e \, \xi \, H_{\rm infl}} \right) \; ,
\end{equation}
is in an excellent agreement with the numerical results, see Fig.~\ref{Bogolyubov}. Here $K_1 \left(x \right)$ with $x \equiv 4(\omega_k (t_e)/a_e) /(\xi H_{\rm infl})$ is the modified Bessel function of the 2nd kind. We restored the scale factor $a_e$ as well as the standard dimensions of $m_{\chi}$ and of the frequency. In the limit $x \ll 1$, one has $K_1(x) \approx \frac{1}{x}$. Consequently, we obtain
\begin{equation}
  \label{small}
  |\beta_k | \simeq \frac{m^2_{\chi} H_{\rm infl}}{24 (\omega_k (t_e)/a_e)^3} \qquad \left(\frac{\omega_k (t_e)}{a_e} \ll \xi H_{\rm infl} \right)\; .
\end{equation}
Note that there is no exponential suppression, as it has been expected. Furthermore, the coefficient $\beta_k$ does not depend on $\xi$ for small frequencies. 
This can be easily understood from Eq.~\eqref{Bogolyubovapprox}, if one neglects the argument in the exponent and makes the rough estimate $|\beta_k| \simeq |\dot{H} (t_1)| \cdot \Delta t_1/m_{\chi}$ (also assume the limit of small wavenumbers).
We use Eqs.~\eqref{halfwidth} and~\eqref{hubblederivative} and end up with the estimate~\eqref{small}. This rough estimate of $|\beta_k |$ shows that $|\beta_k |$ is independent of $\xi$, because so is the product $|\dot{H} (t_1)| \cdot \Delta t_1$: the higher the spike, the narrower it is.

In the opposite limit $x \gg 1$, one has $K_1(x) \approx \sqrt{\frac{\pi}{2}} \cdot \frac{e^{-x}}{\sqrt{x}}$. Hence,
\begin{equation}
  \nonumber
  |\beta_k | \simeq \frac{\sqrt{2\pi}\cdot m^2_{\chi} \cdot H^{1/2}_{\rm infl}}{24\cdot  (\omega_k (t_e)/a_e)^{5/2}} \cdot \mbox{exp} \left[-\frac{4\cdot \omega_k (t_e)}{a_e \cdot \xi \cdot H_{\rm infl}} \right] \qquad \left(\frac{\omega_k (t_e)}{a_e} \gg \xi H_{\rm infl} \right)\; .
\end{equation}
We see explicitly the exponential suppression, which starts at $\omega_k (t_e)/a_e \simeq 4\xi H_{\rm infl}$, in agreement with our expectations.

Two important comments are in order here. The contribution of other spikes does not considerably affect the estimate of the Bogolyubov coefficient $\beta_k$.
The reason is that the contributions of the spikes to $\beta_k$ come with uncorrelated phases. Therefore, these contributions
neither get accumulated nor cancel each other\footnote{Here it is important that $|\beta_k | \ll 1$ in our case. This situation is quite different from the case of parametric resonance, which takes place due to the coupling of the inflaton to a scalar field~\cite{Kofman:1997yn}. In that case, the Bogolyubov coefficient receives an order one contributions at each zero crossing of the inflaton. This leads to the exponential amplification of the number density of produced scalar particles and, consequently to a fast decay of the inflaton.}. Second, recall that our analysis is valid only for small $|\beta_k| \ll 1$.
This limits the range of frequencies $\omega_k$, for which our calculations are applicable. In the regime of interest, $k/a_e \lesssim m_{\chi}$, the condition $|\beta_k| \ll 1$ applied to Eq.~\eqref{small} translates into a lower bound on the mass $m_{\chi}$:
\begin{equation}
  \label{lowermchi}
  m_{\chi} \gg \frac{H_{\rm infl}}{10} \sim H_e \; .
\end{equation}
Hence, the region of applicability of our analysis is limited to masses above the Hubble rate at the end of inflation, which is well enough for our purposes.

The Bogolyubov coefficient is related to the number density of produced particles through
\begin{equation}
  \label{Bogconcentration}
  dn_{\chi}=\frac{k^2 dk}{2\pi^2} \cdot  \frac{1}{a^3(t)} |\beta_k|^2 \; .
\end{equation}
The approximate expression for $n_{\chi} (t)$, which gives the correct analytical formula in the regime of small $m_{\chi}$
and captures the numerically obtained behaviour at large $m_{\chi}$, reads
\begin{equation}
  \nonumber
  n_{\chi} (t) \simeq \frac{ m_{\chi} \cdot H^2_{\rm infl}}{18 \cdot 10^3 \cdot \pi}\cdot F\left(\frac{m_{\chi}}{H_{\rm infl}} \right) \cdot \mbox{exp} \left[-\frac{8 \cdot m_{\chi}}{\xi \cdot H_{\rm infl}} \right]  \cdot  \left(\frac{a_e}{a(t)} \right)^3 \; .
\end{equation}
Here $F\left(\frac{m_{\chi}}{H_{\rm infl}} \right)$ is a fitting function given by
\begin{equation}
  \label{fitting}
  F \left(\frac{m_{\chi}}{H_{\rm infl}} \right)=\frac{1}{\sqrt{1+\frac{m_{\chi}}{2\xi H_{\rm infl}}}} \; .
\end{equation}
Note that for $m_{\chi} \ll \xi H_{\rm infl}/8$, the number density $n_{\chi}$ does not depend on the constant $\xi$. This is a reflection of the analogous property of the coefficient $\beta_k$ discussed above.
The integral over momenta $k$ leading to $n_{\chi}$ is saturated at
$k/a_e \lesssim m_{\chi}$. Hence, one can treat the produced particles as non-relativistic. Thus, the energy density of particles $\chi$ reads
\begin{equation}
  \label{chienergy}
  \rho_{\chi} (t)\simeq \frac{m^2_{\chi} \cdot H^2_{\rm infl}}{18 \cdot 10^3 \cdot \pi} \cdot F\left(\frac{m_{\chi}}{H_{\rm infl}} \right) \cdot  \mbox{exp} \left[-\frac{8 \cdot m_{\chi}}{\xi \cdot H_{\rm infl}} \right]  \cdot \left(\frac{a_e}{a(t)} \right)^3 \; .
\end{equation}
This expression will be the starting point when discussing cosmological applications of $\chi$-particles in the next Sections. Note that the energy density $\rho_{\chi} (t)$ is initially
small relative to the total energy density of the Universe. Consequently, one can neglect the backreaction of the particles $\chi$ on the dynamics of the inflaton. The same is true for inflaton particles created due to the quartic self-interaction: they give a small contribution to the total energy density of the inflaton~\cite{Ema:2016dny}.
On the other hand, light SM particles can be produced abundantly depending on the strength of their couplings to the inflaton and the Ricci scalar. At some point, their backreaction 
on cosmological dynamics may become non-negligible compromising post-inflationary evolution discussed in Section~3 and, consequently, gravitational production of superheavy particles. However, our estimate of the energy density of particles $\chi$ is trustworthy, if this 
backreaction is small enough at the onset of the post-inflationary evolution, so that the shape of the first spikes in Fig.~\ref{fig_Hubble} is not modified considerably. 
It is easy to fulfil this condition by choosing not very large couplings of SM species to the inflaton and gravity. This mild assumption is enough for our purposes. But it is strengthened in Section~5, where gravitational reheating is discussed.

As it follows from Eq.~\eqref{chienergy}, the particles $\chi$ are abundantly created for masses up to $m_{\chi} \sim \xi H_{\rm infl}$. From Eq.~\eqref{constraint}, this would naively mean that one could extend the analysis to the Planck scale $M_{Pl}$ for $\xi \simeq 49000$.
However, we can treat consistently only the masses $m_{\chi}$ below the strong coupling scale~\eqref{strong}.
Particles with masses $m_{\chi}=c \cdot \xi \cdot H_{\rm infl}$, with $c\gtrsim 1$, are created in the weakly coupled regime, provided that the constant $\xi$ is limited as
\begin{equation}
  \label{nostrongnew}
  \xi \lesssim \sqrt{\frac{M_{Pl}}{c  H_{\rm infl}}} \; .
\end{equation}
This is slightly stronger than the bound of Eq.~\eqref{nostrong}.

So far, we have discussed only conformal coupling of particles $\chi$ to gravity. Let us comment on the case $\zeta \neq 1/6$.
We integrate by parts the generic expression~\eqref{Bogreal} valid
in the regime $|\omega'_k/\omega^2_k| \ll 1$:
\begin{equation}
  \label{nonconformal}
  \beta_k \approx -\frac{i}{4} \int^{+\infty}_{t_{Pl}} dt' \frac{\dot{H} (t') \cdot \left[m^2_{\chi} +2(1-6\zeta) \cdot (\omega_k (t')/a(t'))^2 \right] }{(\omega_k (t')/a(t'))^3} \cdot \mbox{exp} \left[-2i \int^{t'}_{t_{Pl}} dt'' \frac{\omega_k (t'')}{a(t'')} \right] \; .
\end{equation}
Here we omitted the terms suppressed by $a \cdot H/\omega_k$ and $a \cdot \dot{\omega_k}/\omega^2_k$.
Note that for $m_{\chi} \ll \xi H_{\rm infl}$ the second term in the square brackets of Eq.~\eqref{nonconformal} gives the dominant contribution to the number density of produced particles (unless $\zeta$ is very close to $1/6$) because the corresponding integral
over momenta is saturated at larger $k/a(t)$, close to $\xi H_{\rm infl}$. Thus, we can take the limit $m_{\chi} \rightarrow 0$ and consider only large $k$ such that $\omega_k/a(t) \approx k/a(t)$.
Then, neglecting the argument in the  exponent and using Eqs.~\eqref{halfwidth} and~\eqref{hubblederivative}, we obtain
\begin{equation}
  \label{BogEma}
  |\beta_k | \simeq \frac{H_{\rm infl} \cdot |1-6\zeta|}{10 \cdot (k/a_e)} \; .
\end{equation}
Modulo the factor two, this estimate is in agreement with the one of Ref.~\cite{Ema:2016dny} (see Eq.~(B.27) there)\footnote{In Eq.~(B.27) of Ref.~\cite{Ema:2016dny}, one should substitute $m_{sp}\simeq (\Delta t_1)^{-1} $ and
$\tilde{m}^2_{\chi} =\frac{1}{6}|1-6\zeta| \cdot |R| \approx |\dot{H}| \cdot |1-6\zeta|$ and use Eqs.~\eqref{halfwidth} and~\eqref{hubblederivative}.}.

Naively, from Eqs.~\eqref{Bogconcentration} and~\eqref{BogEma},
it should follow that the number density of produced $\chi$-particles is enhanced by the factor $\sim 10 \xi |1-6\zeta|^2 H_{\rm infl}/m_\chi$ compared to the case of conformal coupling.
There is, however, another effect, which takes place for large (positive) couplings $\zeta$, such that
\begin{equation}
  \nonumber
  m^2_{\chi} \lesssim \frac{1}{6} |1-6\zeta| \cdot |R|
\end{equation}
at some times in the post-inflationary Universe. The maximal value of $|R|$ is reached at the first spike due to the large value of $\dot{H}$, and the above inequality can be rewritten as
\begin{equation}
  \label{omegapositive}
  m_{\chi} \lesssim \frac{\sqrt{|1-6\zeta| \cdot  \xi}}{5}  \cdot H_{\rm infl} \; .
\end{equation}
If this condition is fulfilled, the field $\chi$ develops a tachyonic instability, because $\omega^2_k$ becomes negative for some $k$. In that regime,
adiabaticity is grossly violated, and particle production may be dramatically amplified. The results of our paper are not applicable to this situation. The reader is referred to Refs.~\cite{Greene:1997ge, Bassett:1997az, Fairbairn:2018bsw} for details.

In the remainder of the paper, we continue assuming conformal coupling to gravity.

\section{Reheating}
\label{section_reheating}

Reheating in inflation~\eqref{inflaction} with the non-minimal coupling has been discussed in Ref.~\cite{Ema:2016dny}, where it was shown that reheating proceeds in a violent way if the field $\varphi$ has gauge interactions with vector fields. Qualitatively the same picture takes place in the model of inflation extended by means of $R^2$-term~\cite{Bezrukov:2019ylq, He:2018mgb}.
See also Refs.~\cite{DeCross:2015uza, DeCross:2016fdz, vandeVis:2020qcp} for an extension to multifield inflation. The earlier work~\cite{Watanabe:2006ku} studied a similar
model~\eqref{inflaction}, but crucially assumes that the inflaton has a large non-zero expectation value. Furthermore, this work does not account for the spike-like features in the inflaton and Hubble evolution.

In this Section, we discuss reheating assuming no direct interactions (in the Jordan frame) of the inflaton with the field $\chi$ and the SM species, which are also assumed to be conformally coupled to gravity. As the masses of SM particles are negligibly small relative to the energy scales of interest, their production is strongly suppressed. The qualitative picture of reheating in this setup is as follows. According to the discussion
of the previous Section, one can efficiently produce non-relativistic matter in the form of a collection of $\chi$-particles\footnote{The possibility of gravitational reheating through the production of superheavy particles has been discussed in Ref.~\cite{Hashiba:2018iff, Haro:2018zdb} in the context of quintessential inflation~\cite{Peebles:1998qn}. However, to achieve an abundant particle production, one must assume very rapid variations of the metric, which is an {\it ad hoc} assumption in this model.}.
As the energy density of non-relativistic matter redshifts as $1/a^3$,
while the energy density of the inflaton in the model~\eqref{modelbasic} evolves as $1/a^4$, the particles $\chi$, if sufficiently stable, come to dominate the energy budget of the Universe. We assume that they have very small, but non-vanishing, couplings to matter fields and thus decay into radiation at some point. Shortly afterwards, reheating takes place.

Before digging into the details, let us make one comment. In Ref.~\cite{Figueroa:2018twl}, it has been argued that reheating in the picture where the inflaton is coupled only to Einstein gravity is in conflict with observations of gravitational waves. The reasoning is as follows. Radiation, which is being produced gravitationally, redshifts faster or at the same rate as
the inflaton energy density, unless the inflaton has a stiff equation of state $w >1/3$. Nevertheless, with the latter assumption gravitational waves are getting strongly blue-shifted, in conflict with BBN and CMB constraints on gravitational waves~\cite{Caprini:2018mtu}. Note, however, that the conclusions of Ref.~\cite{Figueroa:2018twl} are applied only to the gravitational reheating of Ref.~\cite{Ford:1986sy}, which assumes that the inflaton is minimally coupled to gravity. If the inflaton is non-minimally coupled, there is no need for a stiff post-inflationary era, therefore our scenario is in agreement with the non-observation of gravitational waves produced at inflation.

The expression for the energy density of produced $\chi$-particles is given by Eq.~\eqref{chienergy}.
At some point this energy density starts to dominate over the energy density of the inflaton given by
\begin{equation}
  \nonumber
  \rho_\varphi (t) \simeq 3H^2_{e} M^2_{Pl} \cdot \left(\frac{a_e}{a(t)} \right)^4 \; .
\end{equation}
The equality between $\rho_\chi$ and $\rho_\varphi$ occurs at the time $t_*$ defined from
\begin{equation}
  \nonumber
  \frac{a_e}{a_*} \simeq  \frac{1}{54 \cdot 10^3 \cdot \pi} \cdot \left(\frac{H_{\rm infl}}{H_{e}} \right)^2 \cdot \left(\frac{m_{\chi}}{M_{Pl}} \right)^2 \cdot F\left(\frac{m_{\chi}}{H_{\rm infl}} \right) \cdot \mbox{exp} \left[-\frac{8m_{\chi}}{\xi H_{\rm infl}} \right]  \; .
\end{equation}
The Hubble rate $H_*$ is obtained from the Friedmann equation,
\begin{equation}
  \nonumber
  3H^2_* M^2_{Pl} \simeq 2\rho_{\chi, *} \; .
\end{equation}
As a result we have
\begin{equation}
  \label{eq}
  H_* \simeq \frac{H_{\rm infl} }{2 \cdot 10^9 \cdot \pi^2 } \cdot \left(\frac{H_{\rm infl}}{H_e} \right)^3 \cdot \left(\frac{m_{\chi}}{M_{Pl}} \right)^4 \cdot F^2 \left(\frac{m_{\chi}}{H_{\rm infl}} \right) \cdot \mbox{exp} \left[-\frac{16 m_{\chi}}{\xi H_{\rm infl}} \right] \; .
\end{equation}
The Universe is reheated at some time $t_{\rm reh} \gg t_{*}$ almost immediately upon the decay of the particles $\chi$. The subscript `reh' stands for reheating.

The Hubble rate $H_{\rm reh}$ is related to the reheating temperature $T_{\rm reh}$ by
\begin{equation}
  \nonumber
  H_{\rm reh} = \sqrt{\frac{\pi^2 g_* (T_{\rm reh})}{90}} \cdot \frac{T^2_{\rm reh}}{M_{Pl}} \; ,
\end{equation}
where $g_* (T)$ counts the number of ultra-relativistic degrees of freedom. Combining this equation with Eq.~\eqref{hubbleend}, and using Eq.~\eqref{eq}, we can express the reheating temperature as:
\begin{equation}
  \label{intermediate_new}
  T_{\rm reh} \simeq \frac{3 \cdot 10^{5}~\mbox{GeV}}{g^{1/4}_{*} (T_{\rm reh})} \cdot \left(\frac{H_{\rm reh}}{H_*} \right)^{1/2}  \cdot
  \left(\frac{m_{\chi}}{10^{15}~\mbox{GeV}}\right)^2 \cdot F\left(\frac{m_{\chi}}{H_{\rm infl}} \right) \cdot \mbox{exp} \left[-\frac{8m_{\chi}}{\xi H_{\rm infl}} \right] \; .
\end{equation}
Hence, in our scenario the reheating temperatures are relatively low, at least for $H_{\rm reh}/H_{*} \ll 1$.

Let us show that the condition $H_{\rm reh}/H_* \gtrsim 1$ is in conflict with cosmological observations. The reason is that the inflaton plays the role of dark radiation effectively. Thus, were the Universe reheated at times $t \lesssim t_*$, the fraction of dark radiation would be of order unity, which is excluded by studies of BBN (see below). Therefore, the strong inequality
$H_{\rm reh} \ll H_{*}$ should be imposed. Let us quantify this statement using the existing limits on dark radiation. This will also give us the upper bound
on the reheating temperature for a fixed $m_{\chi}$.

First, one defines the fraction of the inflaton energy density relative to the radiation energy density of SM particles at the times of BBN:
\begin{equation}
  \nonumber
  f_{DR} \equiv \frac{\rho_\varphi}{\rho_{rad}}  \left. \right |_{T_{BBN}} \; .
\end{equation}
It is convenient to absorb the effect of dark radiation into the deviation of the effective number of neutrino species from the SM prediction $N_{\nu, SM} \approx 3.046$.
Namely, the fraction $f_{DR}$ can be written as follows:
\begin{equation}
  \label{drneutrino}
  f_{DR} =\frac{7}{4} \cdot \left(\frac{4}{11} \right)^{4/3} \cdot \frac{\Delta N_{\nu}}{g_* (T_{BBN})} \; ,
\end{equation}
where $\Delta N_{\nu}$ denotes the deviation from the SM prediction, i.e., $\Delta N_{\nu} \equiv N_{\nu}-N_{\nu, SM}$; the effective number of the degrees of freedom at the BBN epoch equals $g_*(T_{BBN}) \approx 3.4$. The recent Planck measurement of $N_{\nu}$ reads~\cite{Aghanim:2018eyx}:
\begin{equation}
  \label{planckneutrino}
  N_{\nu}=2.99 \pm 0.17 \; ,
\end{equation}
where the errors are given at $68\%~\mbox{CL}$. That is, the difference $\Delta N_{\nu}$ is limited as $\Delta N_{\nu} \lesssim 0.1$. Then, we express the ratio of the inflaton and $\chi$-particles energy densities through $\Delta N_{\nu}$:
\begin{equation}
  \label{rationeutrino}
  \frac{\rho_\varphi (t_{\rm reh})}{\rho_{\chi} (t_{\rm reh})} \simeq \frac{g^{1/3}_* (T_{\rm reh})}{2 \cdot g^{4/3}_* (T_{BBN})} \cdot \Delta N_{\nu} \; .
\end{equation}
To obtain the above expression, we used an approximate entropy conservation in the comoving volume $s \cdot a^3 \approx \mbox{const}$ and the standard expressions for the energy density and entropy density of radiation:
\begin{equation}
  \label{radentropy}
  \rho_{rad} (T)=\frac{\pi^2 g_* (T) \cdot T^4}{30} \qquad s (T)=\frac{2\pi^2 h_{*} (T)\cdot T^3}{45} \; .
\end{equation}
In Eq.~\eqref{rationeutrino}, we assumed the equality $\rho_{\chi} (t_{reh}) \simeq \rho_{rad} (T_{reh})$ at reheating. We also ignored the inessential difference between the number of ultra-relativistic degrees of freedom $g_* (T)$ and $h_* (T)$
entering energy and entropy densities, respectively. Taking into account the scaling behaviour $\rho_\varphi \propto 1/a^4$ and $\rho_{\chi} \propto 1/a^3$, using $2\rho_\varphi (t_*) \simeq 2\rho_{\chi} (t_*) \simeq 3H^2_*M^2_{Pl}$ and
$2\rho_{\chi} (t_{\rm reh}) \simeq 3H^2_{\rm reh} M^2_{Pl}$, we get
\begin{equation}
  \nonumber
  \frac{\rho_\varphi (t_{\rm reh})}{\rho_{\chi} (t_{\rm reh})} \simeq  \frac{a_*}{a_{\rm reh}} \simeq \left(\frac{H_{\rm reh}}{H_*} \right)^{2/3}\; .
\end{equation}
Then using Eq.~\eqref{rationeutrino} and the previous expression, we obtain
\begin{equation}
  \label{reheq}
  \frac{H_{\rm reh}}{H_*} \simeq \frac{g^{1/2}_* (T_{\rm reh})}{3 \cdot g^{2}_* (T_{BBN})} \cdot \Delta N^{3/2}_{\nu} \; .
\end{equation}
Substituting this expression into Eq.~\eqref{intermediate_new}, one gets
\begin{equation}
  \nonumber
  T_{\rm reh} \simeq \frac{1.7 \cdot \left( \Delta N_{\nu} \right)^{3/4} \cdot 10^{5}~\mbox{GeV}}{g_* (T_{BBN})}  \cdot
  \left(\frac{m_{\chi}}{10^{15}~\mbox{GeV}}\right)^2 \cdot F \left(\frac{m_{\chi}}{H_{\rm infl}} \right)  \cdot \mbox{exp}\left[-\frac{8m_{\chi}}{\xi H_{\rm infl}} \right] \; .
\end{equation}
Using the Planck bounds on the effective number of neutrino species~\eqref{planckneutrino}, we end up with the constraint
\begin{equation}
  \label{upperbound}
  T_{\rm reh} \lesssim  10~\mbox{TeV} \cdot \left(\frac{m_{\chi}}{10^{15}~\mbox{GeV}} \right)^2 \cdot F\left(\frac{m_{\chi}}{H_{\rm infl}}\right) \cdot \mbox{exp}\left[-\frac{8m_{\chi}}{\xi H_{\rm infl}} \right] \; .
\end{equation}
There are prospects of improving the upper bound on $T_{\rm reh}$, albeit not dramatic, through strengthening the constraints on the effective number of neutrino species
in the future meusurements of $N_{\nu}$~\cite{Abazajian:2013oma, Errard:2015cxa, Fields:2019pfx}.

The lower bound on the reheating temperature is set by the requirement of successful BBN~\cite{Hannestad:2004px}:
\begin{equation}
  \label{mibound}
  T_{\rm reh}>4.2~\mbox{MeV}
\end{equation}
at $95\%$~\mbox{CL}. As we can see, the range of masses $H_e \lesssim m_{\chi} \ll \xi H_{\rm infl}$ satisfies the constraints~\eqref{upperbound} and~\eqref{mibound}.
The allowed mass range may change if DM particles produced by the standard freeze-out or freeze-in mechanism are observed in future experiments. Such an observation would considerably increase
the lower bound on $T_{\rm reh}$ and potentially set a lower limit on $m_{\chi}$, or even rule out this model of reheating. Furthermore, a number of models
explaining the baryon asymmetry of the Universe assumes temperatures well above the MeV-range. Keeping this in mind, we continue with the model-independent analysis.

For larger masses $m_{\chi} \gtrsim \xi H_{\rm infl}$, the results are modified due to the exponential suppression of the energy density of produced particles. Consequently, it takes more time before the $\chi$-particles start dominating evolution of the Universe.
As a result, typical reheating temperatures turn out to be very low. In particular, the temperature $T_{\rm reh} \simeq 4~\mbox{MeV}$ is reached for a mass $m_{\chi}$ satisfying the equation
\begin{equation}
  \nonumber
  \frac{8m_{\chi}}{\xi H_{\rm infl}} \approx 9+2\ln \xi +\frac{3}{2}\ln \frac{m_{\chi}}{\xi H_{\rm infl}} \; ,
\end{equation}
where we used Eqs.~\eqref{fitting} and (\ref{upperbound}). For $\xi =100$, the above equation yields the constraint on the mass $m_{\chi}$:
\begin{equation}
  \nonumber
  m_{\chi} \lesssim 2.5 \cdot \xi H_{\rm infl} \; .
\end{equation}
The upper limit here is not altered significantly for different $\xi$. Note that $\chi$-particles with even larger masses are still of interest in the cosmological context: they can constitute DM, if stable. We consider this option in the next Section.

Before that, let us discuss the strength of couplings of the particles $\chi$ to other matter fields. We assume that the Universe is mainly reheated
due to Yukawa coupling of the field $\chi$ to fermions $S$ (e.g., sterile neutrinos), with the interaction Lagrangian
\begin{equation}
  \label{yukawa}
  {\cal L}_{int}=y \chi \bar{S} S \; .
\end{equation}
The fermions $S$ subsequently decay into SM species. The Yukawa coupling $y$ can be found from
\begin{equation}
  \nonumber
  \Gamma_{\chi \rightarrow S} \approx \frac{y^2 m_{\chi}}{8\pi} \simeq H_{\rm reh}  \; .
\end{equation}
Using Eqs.~\eqref{eq},~\eqref{reheq}, and the Planck bounds~\eqref{planckneutrino}, and substituting $g_*(T_{\rm reh}) \simeq 100$, one gets the upper limit on the Yukawa coupling:
\begin{equation}
  \nonumber
  y \lesssim 1.6 \cdot 10^{-12}  \cdot \left(\frac{m_{\chi}}{10^{15}~\mbox{GeV}} \right)^{3/2} \cdot F \left(\frac{m_{\chi}}{H_{\rm infl}} \right) \cdot
  \exp \left[-\frac{8 m_{\chi}}{ \xi H_{\rm infl}} \right] \; .
\end{equation}
Therefore, the coupling of the field $\chi$ to matter fields must be very weak. For the Yukawa coupling close to the upper bound, the
fraction of dark radiation can be potentially testable with the future measurements of the effective number of neutrino species. The lower bound
on $y$ follows from the requirement that the decay $\chi \rightarrow S$ occurs before the temperature $T \simeq 4~\mbox{MeV}$ is reached, $\Gamma_{\chi \rightarrow S} \gtrsim H \left. \right|_{T \simeq 4~MeV}$:
\begin{equation}
 \nonumber
  y \gtrsim 4 \cdot 10^{-19} \cdot \sqrt{\frac{10^{15}~\mbox{GeV}}{m_{\chi}}} \; .
\end{equation}
We see that even for the masses $m_{\chi} \simeq 10^{13}~\mbox{GeV}$, the region of allowed values of $y$ spans three orders of magnitude.

\section{Dark Matter}
\label{section_dark_matter}

As shown in the previous Sections, particles $\chi$ are abundantly created in the mass range $H_e \lesssim m_{\chi} \ll\xi H_{\rm infl}$.
If they were stable, they would overclose the Universe well before the conventional matter-radiation equality. Hence, particles $\chi$ in this mass range cannot be considered for the role of DM.
A way out of this problem is to assume that the $\chi$-particles decay into lighter stable particles. The latter can play the role of DM in a certain range of parameters. 
Another option is to consider heavier masses, $m_{\chi} \gtrsim \xi H_{\rm infl}$. In this case, the number density of produced particles $\chi$ is exponentially suppressed.
Then, upon a proper choice of model parameters, one can adjust the suppression and achieve the right abundance of the particles $\chi$, so that they can constitute DM.
In this Section, we consider both options and assume that the inflaton has direct interactions with the matter fields, apart from $\chi$-particles, so that the Universe gets quickly reheated in the standard fashion.

\subsection{Particles $\chi$ as Dark Matter}
\label{section_chi_as_dark_matter}

Let us first consider the case of very heavy particles $\chi$ with the masses
\begin{equation}
  \label{supersuper}
  m_{\chi} \gtrsim \xi H_{\rm infl} \; .
\end{equation}
We are interested in the scenario, when particles $\chi$ constitute all DM in the Universe. Hence, the following condition should be obeyed:
\begin{equation}
  \label{radmatequality}
  \rho_{rad} (t_{eq}) \approx \rho_{\chi} (t_{eq}) \; ,
\end{equation}
where the subscript $'eq'$ stands for the matter-radiation equality; the energy density of radiation is given by Eq.~\eqref{radentropy}.
The energy density $\rho_{\chi} (t_{eq})$ can be found from Eq.~(\ref{chienergy}), where one sets $t=t_{eq}$. We decompose the ratio $(a_e/a_{eq})^3$ in Eq.~(\ref{chienergy}) as follows:
\begin{equation}
  \label{ratioeeq}
  \left(\frac{a_e}{a_{eq}} \right)^3=\left(\frac{a_e}{a_{\rm reh}} \right)^3 \cdot \left(\frac{a_{\rm reh}}{a_{eq}} \right)^3 \; .
\end{equation}
The ratio $(a_e/a_{\rm reh})^3$ is given by
\begin{equation}
  \label{ratioereh}
  \left(\frac{a_e}{a_{\rm reh}} \right)^3 \simeq \left(\frac{\pi^2 g_{*} (T_{\rm reh}) \cdot T^4_{\rm reh}}{90H^2_e M^2_{Pl}} \right)^{\frac{3}{4}} \; .
\end{equation}
Here we take into account that the post-inflationary evolution until reheating is described by the radiation-like equation of state,
which is characteristic for the inflationary model with the quartic potential. The ratio $(a_{\rm reh}/a_{eq})^3$ is inferred from the approximate entropy conservation in the comoving volume (see Eq.~\eqref{radentropy}):
\begin{equation}
  \label{ratioreheq}
  \left(\frac{a_{\rm reh}}{a_{eq}} \right)^3 \approx \frac{T^3_{eq} \cdot g_{*} (T_{eq})}{T^3_{\rm reh} \cdot g_{*} (T_{\rm reh})} \; .
\end{equation}
We again ignore the difference between the numbers $g_{*} (T)$ and $h_* (T)$.
Combining the above expressions, using Eqs.~\eqref{hubbleend},~\eqref{chienergy}, and~\eqref{radentropy}, and substituting $g_{*} (T_{\rm reh}) \simeq 100$, we obtain the equation, which determines the mass $m_{\chi}$:
\begin{equation}
  \nonumber
  2 \cdot 10^{11} \cdot \xi^2 \cdot \left(\frac{m_{\chi}}{\xi H_{\rm infl}} \right)^2 \cdot F\left(\frac{m_{\chi}}{H_{\rm infl}} \right) \cdot \mbox{exp} \left[-\frac{8m_{\chi}}{\xi H_{\rm infl}} \right]\simeq 1 \; .
\end{equation}
Taking the logarithm of the latter, we get
\begin{equation}
  \nonumber
  \frac{8 m_{\chi}}{\xi H_{\rm infl}} \approx 29+2 \ln \xi +\frac{3}{2}\ln \frac{m_{\chi}}{\xi H_{\rm infl}} \; .
\end{equation}
For $\xi = 100$, the solution of the above equation is
\begin{equation}
  \nonumber
  m_{\chi} \approx 5 \cdot \xi H_{\rm infl} \; .
\end{equation}
As it follows from Eq.~\eqref{nostrongnew}, these large masses can be treated consistently, provided that the constant $\xi$ is limited as $\xi \lesssim 100$.
Otherwise, the mass $m_{\chi}$ exceeds the strong coupling scale~\eqref{strong}, and we cannot trust our analysis. We conclude with the following constraint on the mass $m_{\chi}$:
\begin{equation}
  \label{upperdark}
  m_{\chi} \lesssim 3 \cdot 10^{16}~\mbox{GeV} \; ,
\end{equation}
where the upper bound is reached for $\xi \simeq 100$.

Another approach for particle creation, which avoids strong coupling issue is to consider an extension of the inflationary model~(\ref{inflaction}) by introducing the $R^2$-term~\cite{Ema:2017rqn, Gorbunov:2018llf}. In this extension, for a suitable
choice of parameters, the strong coupling takes place only at the Planck scale. At the same time, however, the spikes become smoother.
Therefore, we do not expect a considerable relaxation of the constraint~\eqref{upperdark} in the extended version.

\subsection{Decay products of particles $\chi$ as Dark Matter}
\label{section_decay_products}

Finally, let us assume that the particles $\chi$ have the masses $m_{\chi} \ll \xi H_{\rm infl}$, but they are unstable and decay into stable fermions
$S$ with the masses $m_S$ through the Yukawa interaction of the form~\eqref{yukawa}. Below we discuss the bounds on the masses $m_S$ and Yukawa coupling constants $y$, which
yield the right abundance of cold DM composed of the particles $S$.
The energy density of $S$-particles at the times, when they have become
non-relativistic, is easily inferred from Eq.~\eqref{chienergy}:
\begin{equation}
  \nonumber
  \rho_S (t) \simeq \frac{m_S \cdot m_{\chi} \cdot H^2_{\rm infl}}{9 \cdot 10^3 \cdot \pi} \cdot \left(\frac{a(t_e)}{a(t)} \right)^3 \; .
\end{equation}
Then, following the same arguments as in the previous Subsection, we obtain
\begin{equation}
  \label{lower}
  m_S \simeq 6~\mbox{GeV} \cdot \left(\frac{10^{15}~\mbox{GeV}}{m_{\chi}} \right) \; .
\end{equation}
The particles $m_{\chi}$ are produced without the exponential suppression in the range of masses $10^{13}~\mbox{GeV} \lesssim m_{\chi} \lesssim 10^{16}~\mbox{GeV}$ for $\xi \simeq 1-100$, where the lower bound comes from Eq.~(\ref{lowermchi}).
This range translates into the range of masses of particles $S$, i.e., $1~\mbox{GeV} \lesssim m_S \lesssim 1~\mbox{TeV}$,
which could be of interest from the viewpoint of ground based experimental searches for DM.

Note that the particles $S$ must be non-relativistic, when the temperature of the Universe is about $1~\mbox{keV}$. Otherwise, one risks to compromise
a well-established bottom-up picture of the structure formation. Hence, the field $\chi$ must decay before the temperature drops down to
\begin{equation}
  \nonumber
  T_{{dec}} \simeq \frac{g^{1/3}_* (T \simeq 1~\mbox{keV})}{g^{1/3}_*(T_{dec})} \cdot \left(\frac{m_{\chi}}{2m_S} \right) \cdot \mbox{keV} \; .
\end{equation}
To paraphrase, the following inequality should be obeyed:
\begin{equation}
  \nonumber
  \Gamma \simeq \frac{y^2 m_{\chi}}{8\pi} \gtrsim H(T_{dec}) \; .
\end{equation}
This yields the lower bound on the allowed value of the coupling constant $y$:
\begin{equation}
  \nonumber
  \frac{y^2}{8\pi} \gg  10^{-18} \cdot \frac{m_{\chi}}{10^{15}~\mbox{GeV}} \cdot \left(\frac{6~\mbox{GeV}}{m_S} \right)^2 \; ,
\end{equation}
where we used $g_* (T_{dec}) \simeq 100$ and $g_* (T \simeq 1~\mbox{keV}) \simeq 3.4$.
For the relevant values of $m_{\chi}$ and $m_S$ a fairly broad range of the coupling $y$ values is allowed.
For values $y$ hitting the lower bound, DM is warm.
This is despite the fact that the masses of $S$-particles are considerably heavier than the canonical value $1-10$~\mbox{keV}.
There is no contradiction, however, because the range $1-10$~\mbox{keV} is obtained for warm DM produced by particles in plasma, while $S$-particles are created by the source ($\chi$-particles) being out of thermal equilibrium.

Let us comment on how the parameter space is altered in this scenario, if the particles $\chi$ have heavier masses $m_{\chi} \gtrsim \xi H_{\rm infl}$ and/or the branching ratio of the decay into the particles $S$ is small
(that is, particles $\chi$ have more dominant decay channels). As a result, the number density of produced particles $S$ for a given mass $m_{\chi}$ is going to be considerably smaller
compared to the case discussed above. Hence, to compensate this and get the right abundance of DM, one should assume larger masses $m_S$. We conclude that Eq.~\eqref{lower} gives a lower bound on the masses $m_{S}$.

\section{Summary}
\label{section_summary}

In the present work, we showed that inflationary scenarios with a non-minimal coupling of the inflaton $\varphi$ to the Ricci curvature, i.e., $\xi \varphi^2 R/2$, provide a perfect playground for gravitational creation of particles with super-Hubble masses.
This was achieved by evaluating the energy density of heavy particles $\chi$ assumed to be conformally coupled to gravity.
We demonstrated that the particles $\chi$ with masses up to the Grand Unification scale can be abundantly created for $\xi \simeq 100$, thanks
to the spike-like behaviour of the time derivative of the Hubble rate.
In the presence of these spikes, particle production proceeds without the exponential suppression for masses $m_{\chi} \lesssim \xi H_{\rm infl}$. Furthermore, for these masses the number density of particles $\chi$ is independent of the coupling constant $\xi$ for $\xi\gg 1$.
The energy density of particles $\chi$ was evaluated both numerically and analytically.
To undertake the latter task, we calculated the inflaton and Hubble rate profiles in the vicinity of the spikes, which allowed us to find the analytical expression for the
Bogolyubov coefficient.

We considered different cosmological scenarios involving the $\chi$ particles. 
We showed that the particles $\chi$ can reheat the Universe if they have direct couplings to SM particles (or sterile neutrinos), even if the inflaton has only gravitational interactions with the other matter fields and decays only due to the cosmic expansion.
This scenario is possible because the energy density of non-relativistic $\chi$-particles redshifts more slowly than the energy density of the inflaton field. After the particles $\chi$ come
to dominate evolution of the Universe, they decay into the SM species, and reheating takes place.
The resulting reheating temperature $T_{\rm reh}$ strongly depends on the
mass $m_{\chi}$, but generally it is low relative to standard inflationary predictions. For example, the upper limit on $T_{\rm reh}$ is about $10~\mbox{TeV}$ for $m_{\chi} \simeq 10^{15}~\mbox{GeV}$ and $\xi \simeq 100$. 
The upper bound on the reheating temperature is stronger for masses both smaller and larger than $m_{\chi} \simeq 10^{15}~\mbox{GeV}$.
The upper bound on $T_{\rm reh}$ comes from the fact that the oscillating inflaton condensate manifests as dark radiation cosmologically.
For some range of model parameters, the fraction of dark radiation in the total radiation can be sizeable and probed through the
measurements of the effective number of neutrino species $N_{\nu}$. The current constraint on $N_{\nu}$ translates into the upper bound on $T_{\rm reh}$.

We also considered applications of particles $\chi$ for DM. There are at least two options. The first and most economical one is to assume that the particles $\chi$ are stable and constitute all DM. In this case, they should be extremely heavy, with masses $m_{\chi} \gtrsim \xi H_{\rm infl}$. The reason is that lighter particles with $m_{\chi} \lesssim \xi H_{\rm infl}$ are overproduced in this scenario, so that they would overclose the Universe.
On the other hand, the production of particles with masses $m_{\chi} \gtrsim \xi H_{\rm infl}$ is exponentially suppressed.
The exponential suppression makes it possible to avoid overproduction of DM, so that the matter-radiation equality constraint at $T_{eq} \simeq 1~\mbox{eV}$ is satisfied. 
For lighter particles $m_{\chi} \lesssim \xi H_{\rm infl}$, a way to avoid overproduction is to consider DM as the product of the decay of $\chi$-particles. For example, one can assume that the latter have Yukawa couplings with stable sterile fermions $S$, which may constitute all DM in a certain range of the parameter space. Specifically, the masses of particles $S$ are constrained to be in the range $1~\mbox{GeV}-1~\mbox{TeV}$.

\section*{Acknowledgments} We are indebted to Sergei Winitzki for useful discussions.
The work of E.B. is supported by the CNRS/RFBR Cooperation program for 2018-2020 n.~1985 ``Modified gravity and black holes: consistent models and experimental signatures''.
The work of D.G. is supported by the Russian Foundation for Basic Research grant 18-52-15001-NCNIa.
S.R. is supported by the European Regional Development Fund (ESIF/ERDF) and the Czech Ministry of Education, Youth and Sports (M\v SMT) through the Project CoGraDS-CZ.02.1.01/0.0/0.0/15 003/0000437. The work of L.R. is supported by the Czech Science Foundation GA\v CR, project 20-16531Y.

\appendix

\numberwithin{equation}{section}

\section{Details of post-inflationary evolution of the inflaton and the Hubble rate}
\label{appendix_details_post_inflat}

In this Appendix, we solve analytically Eqs.~\eqref{friedmanndim} and~\eqref{inflatondim} for the
Hubble rate and the inflaton, respectively, in different regimes.
These equations are written in dimensionless variables introduced in Eq.~\eqref{dimensionless}.
We analyse the system in two regimes of interest: (i) during inflation and (ii) shortly after, when the first spike appears. As in the main body of the paper, we assume large $\xi$, formally $\xi \rightarrow \infty$.
Our discussion below matches that of Ref.~\cite{Ema:2016dny}, whether there is an overlap. Compared to Ref.~\cite{Ema:2016dny}, however, we find the analytical expressions for the
inflaton and the Hubble rate in the vicinity of the first spike. See also Ref.~\cite{DeCross:2016fdz} for the analogous calculations in the Einstein frame.
The results of this Appendix will be the starting point for calculation of the Bogolyubov coefficient in Appendix~\ref{appendix_analytic_bogolyubov}.

During inflation, when $\tilde{\varphi} \gg 1$, the system of equations~\eqref{friedmanndim} and~\eqref{inflatondim} simplifies to
\begin{equation}
  \label{friedmanninflapp}
  3 \left(\tilde{H}^2+\tilde{\varphi}^2 \tilde{H}^2 +2 \tilde{\varphi} \dot{\tilde{\varphi}} \tilde{H} \right)=\frac{\tilde{\varphi}^4}{4}
\end{equation}
and
\begin{equation}
  \label{inflinflapp}
  \tilde{\varphi}^2-6 \left(2\tilde{H}^2+\dot{\tilde{H}}\right)=0 \; .
\end{equation}
The terms omitted vanish in the limit $\xi \rightarrow \infty$. In the zeroth order approximation, one obtains
\begin{equation}
  \label{zerothorder}
  \tilde{\varphi}^2 =12 \tilde{H}^2 \; .
\end{equation}
We are interested in finding values of the Hubble rate and the inflaton as well as their derivatives at the end of inflation.
These can be derived in the next-to-leading order.
For this purpose, it is convenient to rewrite Eqs.~\eqref{friedmanninflapp} and~\eqref{inflinflapp} as follows:
\begin{equation}
  \nonumber
  3 \left(\tilde{H}^2+2\tilde{\varphi} \dot{\tilde{\varphi}} \tilde{H}  \right)=\frac{\tilde{\varphi}^2}{4} \cdot \left[\tilde{\varphi}^2-12 \tilde{H}^2 \right]
\end{equation}
and
\begin{equation}
  \nonumber
  \dot{\tilde{H}}=\frac{1}{6} \cdot \left[\tilde{\varphi}^2-12 \tilde{H}^2 \right] \; .
\end{equation}
Combining the latter two, we obtain
\begin{equation}
  \nonumber
  \dot{\tilde{H}}=2\frac{\tilde{H}^2}{\tilde{\varphi}^2}+4 \frac{\dot{\tilde{\varphi}}}{\tilde{\varphi}} \tilde{H} \; .
\end{equation}
Now using the result~\eqref{zerothorder}, one gets
\begin{equation}
  \label{hubbledirapp}
  \dot{\tilde{H}}=-\frac{1}{18} \; .
\end{equation}
In particular, this can be used to find the Hubble rate at the end of inflation, which occurs roughly when $2\tilde{H}^2_e \sim  |\dot{\tilde{H}}_e|$. We obtain
\begin{equation}
  \label{hubbleendapp}
  \tilde{H}_e \simeq \frac{1}{6} \; .
\end{equation}
The above result is used in the estimate~\eqref{hubbleend} in the main body of the text.
Note that the resulting values of the inflaton and its derivative at the end of inflation
immediately follow from Eqs.~\eqref{zerothorder} and~\eqref{hubbleendapp}: $\tilde{\varphi}_e \simeq 1/\sqrt{3}$ and $\dot{\tilde{\varphi}}_e \simeq -1/(3\sqrt{3})$.

Now, let us switch to the evolution after the end of inflation with the focus on the region around the first spike.
The value of the inflaton field there $\tilde{\varphi}$ approaches zero, while its derivative and consequently the derivative
of the Hubble rate are large. Thus, the system of Eqs.~\eqref{friedmanndim} and~\eqref{inflatondim} can be simplified to
\begin{equation}
  \label{hubbleapp}
  3 \left(\tilde{H}^2+2\tilde{H} \tilde{\varphi} \dot{\tilde{\varphi}} \right)=\frac{1}{2\xi} \dot{\tilde{\varphi}}^2
\end{equation}
and
\begin{equation}
  \label{inflatonapp}
  \ddot{\tilde{\varphi}}-6\xi \tilde{\varphi} \dot{\tilde{H}}=0 \; .
\end{equation}
The Hubble rate is easily expressed from Eq.~\eqref{hubbleapp}:
\begin{equation}
  \label{hubbleanalytic}
  \tilde{H}=-\dot{\tilde{\varphi}} \cdot \left(\tilde{\varphi}+\sqrt{\tilde{\varphi}^2+\frac{1}{6\xi}} \right) \; .
\end{equation}
Substituting this expression into Eq.~\eqref{inflatonapp}, we get
\begin{equation}
  \nonumber
  \ddot{\tilde{\varphi}} \cdot \left(\tilde{\varphi}^2+\frac{1}{6\xi} \right)+\tilde{\varphi}\dot{\tilde{\varphi}}^2 =0 \; .
\end{equation}
The solution reads
\begin{equation}
  \label{genioussolution}
  \tilde{t} (\tilde{\varphi})=C \cdot \left[\tilde{\varphi} \cdot \sqrt{1+6\xi \tilde{\varphi}^2}+\frac{1}{\sqrt{6\xi}} \ln \left(\sqrt{6\xi} \tilde{\varphi}+\sqrt{1+6\xi \tilde{\varphi}^2} \right) \right] \; .
\end{equation}
The second constant of integration was chosen so that $\tilde{t} (0)=0$ without loss of generality.
From the solution~\eqref{genioussolution} it is evident that the spike is occuping the region in the field space:
\begin{equation}
  \nonumber
  -\frac{1}{\sqrt{6\xi}}\lesssim \tilde{\varphi} \lesssim \frac{1}{\sqrt{6\xi}} \; .
\end{equation}
The constant $C$ is obtained from matching the solution~\eqref{genioussolution} to the
behaviour of the inflaton at the end of inflation. To do this, let us take the derivative of Eq.~\eqref{genioussolution} with respect to $\tilde{\varphi}$ at the point $\tilde{\varphi}=\tilde{\varphi}_e$.
Using Eq.~\eqref{hubbleanalytic}, one gets
\begin{equation}
  \label{const}
  C \simeq \frac{1}{\sqrt{6\xi} \tilde{H}_e} \simeq \sqrt{\frac{6}{\xi}} \; .
\end{equation}
Now we can calculate $\dot{\tilde{H}}$ at the center of the spike:
\begin{equation}
  \nonumber
  \dot{\tilde{H}} (0) \approx -\dot{\tilde{\varphi}}^2 (0) \; .
\end{equation}
The value $\dot{\tilde{\varphi}} (0)$ is related to the constant $C$ by $C=1/(2\dot{\tilde{\varphi}}(0))$. Combining the latter and Eq.~\eqref{const}, one obtains
\begin{equation}
  \label{analyticalfinal}
  \dot{\tilde{H}} (0) \simeq -\frac{\xi}{24} \; ,
\end{equation}
which is equivalent to the expression~\eqref{hubblederivative} in the main body of the text.
Finally, let us estimate the width of the spike. This is given by
\begin{equation}
  \label{widthapp}
  \tilde{t} \left(\frac{1}{\sqrt{6\xi}} \right)-\tilde{t} \left(-\frac{1}{\sqrt{6\xi}}\right) \simeq \frac{5}{\xi} \; ,
\end{equation}
which is the value we used in the estimate~\eqref{halfwidth}.

\section{Analytic estimation of Bogolyubov coefficient}
\label{appendix_analytic_bogolyubov}

Using the results of Appendix~\ref{appendix_details_post_inflat}, one can obtain the analytical estimate of the Bogolyubov coefficient $\beta_k$. We start with the expression~\eqref{Bogolyubovapprox}. It is convenient to replace the integration over the time by the integral over the field $\tilde{\varphi}$ (we continue to work with dimensionless variables of Eq.~\eqref{dimensionless}). In particular, we make the following change in the integrand
\begin{equation}
  \nonumber
  d \tilde{t} \cdot  \frac{d\tilde{H}}{d\tilde{t}} =d\tilde{\varphi} \cdot  \frac{d\tilde{H}}{d\tilde{\varphi}} \; .
\end{equation}
We again assume that the main contribution to the Bogolyubov coefficient comes from the vicinity of the first spike. In this region, $\frac{d\tilde{H}}{d\tilde{\varphi}}$ reads
\begin{equation}
  \nonumber
  \frac{d\tilde{H}}{d\tilde{\varphi}}=-\frac{\sqrt{\xi}}{2\sqrt{6} \cdot \left(1+6\xi \tilde{\varphi}^2 \right)^{3/2}} \; ,
\end{equation}
which follows from Eqs.~\eqref{hubbleanalytic},~\eqref{genioussolution}, and~\eqref{const}. Furthermore, at $|\tilde{\varphi}| \ll 1$, one can approximate Eq.~\eqref{genioussolution} by $\tilde{t} \approx 2\sqrt{\frac{6}{\xi}} \cdot \tilde{\varphi}$, where we used Eq.~\eqref{const}. As in Appendix~\ref{appendix_details_post_inflat}, we set $\tilde{t}=0$ at the center of the spike, where $\tilde{\varphi}=0$. Hence, the integral of interest is given by
\begin{equation}
  \label{Bogint}
  \beta_k \simeq  \frac{i\cdot \tilde{m}^2_{\chi} \cdot \sqrt{\xi}}{8 \cdot \sqrt{6} \cdot \tilde{\omega}^3_k} \int^{+\infty}_{-\infty}  \frac{d \tilde{\varphi}}{ \left(1+6\xi \tilde{\varphi}^2 \right)^{3/2}} \cdot \mbox{exp}
  \left[-4i \sqrt{\frac{6}{\xi}} \tilde{\omega_k} \tilde{\varphi} \right] \; ,
\end{equation}
where
\begin{equation}
  \nonumber
  \tilde{m}_{\chi} \equiv \frac{m_{\chi}}{H_{\rm infl}} \qquad \tilde{\omega}_k \equiv \frac{\omega_k}{H_{\rm infl}} \; .
\end{equation}
Note that we keep the scale factor constant around the spike and normalize it to unity, i.e., $a(\tilde{t}) \approx a(0) =1$. This is legitimate, because the time scale of the spike is very short, so that the scale factor
has no time to change considerably. Upon the change of a variable, $z=\sqrt{6\xi}\tilde{\varphi}$, Eq.~\eqref{Bogint} takes the form:
\begin{equation}
  \nonumber
  \beta_k =\frac{i\tilde{m}^2_{\chi}}{48 \cdot \tilde{\omega}^3_k} \int^{+\infty}_{-\infty} \frac{dz}{(1+z^2)^{3/2}} \cdot \mbox{exp} \left[-\frac{4i}{\xi}\tilde{\omega}_k z \right] \; .
\end{equation}
To proceed, we make use of the integral representation of the modified Bessel function of the 2nd kind:
\begin{equation}
  \nonumber
  K_1 (x)=\frac{1}{x} \int^{+\infty}_0 \frac{dz}{(1+z^2)^{3/2}} \cdot \cos (x \cdot z) \; .
\end{equation}
Comparing the latter two expressions, we obtain the analytic estimate of the Bogolyubov coefficient:
\begin{equation}
  \label{Bogolyubovanalytic}
  \beta_k \simeq \frac{i \cdot \tilde{m}^2_{\chi}}{6\cdot \tilde{\omega}^2_k \cdot \xi} \cdot K_1 \left(\frac{4\tilde{\omega}_k}{\xi} \right) \; .
\end{equation}
This is in an excellent agreement with the numerical results, see Fig.~\eqref{Bogolyubov}.

\section{On numerical calculation of the Bogolyubov coefficient}
\label{appendix_numerical_bogolyubov}

We start with the expression~\eqref{Bogolyubovconformal} for the Bogolyubov coefficient $\beta_k$, which we repeat here for convenience of references:
\begin{equation}
  \label{repeat}
  \beta_k =\frac{1}{2} \int^{+\infty}_{t_{Pl}} dt' \cdot \frac{m^2_{\chi} \cdot H(t')}{(\omega_k (t')/a(t'))^2}  \cdot \mbox{exp} \left[-2i \int^{t'}_{t_{Pl}} dt'' \frac{\omega_k (t'')}{a(t'')} \right] \; .
\end{equation}
The above expression is not very convenient for the purpose of numerical calculations, because the pre-exponential function in the integrand is non-vanishing during inflation.
While the integral remains converging for any $k \neq 0$, because the scale factor $a(t) \rightarrow 0$ at very early times, the exponent is oscillating fast in this limit. This complicates numerical calculations.
The same obstacle occurs for the expression~\eqref{intermediate_new}. Indeed, the derivative $\dot{H}$ does not turn into zero during inflation--it remains a constant, albeit small, see Eq.~\eqref{hubbledirapp}.

Therefore, we perform two integrations by parts in Eq.~\eqref{repeat} (or one integration by parts in Eq.~\eqref{intermediate_new}) and obtain
\begin{equation}
  \begin{split}
    \beta_k &=-\frac{1}{8} \int^{+\infty}_{t_{Pl}} dt' m^2_{\chi} \cdot \Bigl[\frac{\ddot{H} (t')}{(\omega_k (t')/a(t'))^4}+\frac{10 H(t') \cdot \dot{H} (t')}{(\omega_k (t')/a(t'))^6} \cdot \frac{k^2}{a^2(t')}+
    \\ &+\frac{12H^3(t')}{(\omega_k (t')/a(t'))^8}
    \cdot \frac{k^4}{a^4(t')}-\frac{6m^2_{\chi} H^3(t')}{(\omega_k (t')/a(t'))^8} \cdot \frac{k^2}{a^2(t')} \Bigr] \cdot \mbox{exp} \left[-2i \int^{t'}_{t_{Pl}} dt'' \frac{\omega_k (t'')}{a(t'')} \right] \; .
  \end{split}
\end{equation}
This is an exact expression. To simplify it, we first observe that the contributions to the Bogolyubov coefficient due to the third and the fourth terms
in the square brackets are suppressed compared to the basic expression~\eqref{repeat} by the factor $ \sim H^2/\omega^2_k \ll 1$. Hence, they can be safely dropped.
The same is less evident for the second term. To show this, let us replace $H \dot{H}$ by $dH^2/2dt$ and then perform the integration by parts. Namely,
\begin{equation}
  \begin{split}
    \beta_k \supset & -\frac{5}{8} \int^{+\infty}_{t_{Pl}} dt' ~\frac{d H^2(t')}{dt'} \cdot \frac{m^2_{\chi} \cdot (k^2/a^2(t'))}{(\omega_k (t')/a(t'))^6} \cdot \mbox{exp} \left[-2i \int^{t'}_{t_{Pl}} dt'' \frac{\omega_k (t'')}{a(t'')} \right] \approx \\
    & \approx -\frac{5i}{4} \int^{+\infty}_{t_{Pl}} dt' ~\frac{m^2_{\chi} \cdot H^2 (t') \cdot (k^2/a^2(t'))}{(\omega_k (t')/a(t'))^5} \cdot \mbox{exp} \left[-2i \int^{t'}_{t_{Pl}} dt'' \frac{\omega_k (t'')}{a(t'')} \right] \; .
  \end{split}
\end{equation}
It is clear that the latter expression is suppressed compared to the one of Eq.~\eqref{repeat} by the factor $H/\omega_k$.

Consequently, the expression for the Bogolyubov coefficient simplifies to
\begin{equation}
  \nonumber
  \beta_k \approx -\frac{1}{8} \int^{+\infty}_{t_{Pl}} dt' ~\frac{m^2_{\chi} \cdot \ddot{H} (t')}{(\omega_k (t')/a(t'))^4} \cdot \left[-2i \int^{t'}_{t_{Pl}} dt'' \frac{\omega_k (t'')}{a(t'')} \right] \; .
\end{equation}
This expression is more suitable for numerical calculations, because the second derivative $\ddot{H}$ tends to zero during inflation. Finally, assuming that the scale factor
does not change considerably within the first spike, which gives the dominant contribution to the Bogolyubov coefficient, we can write
\begin{equation}
  \nonumber
  \beta_k =-\frac{1}{8} \frac{m^2_{\chi}}{(\omega_k (t_e)/a_e)^4} \cdot \int^{+\infty}_{t_{Pl}} \ddot{H} (t') \cdot \mbox{exp} \left[-2i \frac{\omega_k (t_e) \cdot t'}{a_e} \right] \; .
\end{equation}
This is the expression we deal with when performing numerical calculations. The result is in a very good agreement with the analytical estimate~\eqref{Bogolyubovanalytic}.

\end{document}